\documentstyle[sprocl,epsfig]{article}

\def\Journal#1#2#3#4{{#1} {\bf #2}, #3 (#4)}

\def\NPB{{\em Nucl. Phys.} B}
\def\PLB{{\em Phys. Lett.}  B}
\def\PRL{\em Phys. Rev. Lett.}
\def\PRD{{\em Phys. Rev.} D}

\def\be{\begin{equation}}
\def\ee{\end{equation}}
\def\bea{\begin{eqnarray}}
\def\eea{\end{eqnarray}}

\begin{document}


\setcounter{page}{0}
\pagestyle{empty}

\begin{flushright}
HIP-2000-57/TH 
\end{flushright}
\vspace*{26mm}

\title{QCD PHASE TRANSITION AND\\  
       PRIMORDIAL DENSITY 
PERTURBATIONS$^*$\protect\footnotemark[0] \vspace*{8mm} }
\footnotetext[0]{$^*$
 To appear in the Proceedings of COSMO-2000, 4th International Workshop
 on Particle Physics and the Early Universe,
 Cheju Island, South Korea, 4--8 September 2000.}

\author{J. IGNATIUS \vspace*{1mm} }

\address{Department of Physics,
         P.O. Box 9, FIN-00014 University of Helsinki, Finland\\
E-mail: janne.ignatius@iki.fi \vspace*{3mm} }

\author{DOMINIK J. SCHWARZ \vspace*{1mm} }

\address{Institut f\"ur Theoretische Physik, 
         TU Wien,\\ 
         Wiedner Hauptstra\ss e 8 -- 10, A-1040 Wien, Austria\\ 
E-mail: dschwarz@hep.itp.tuwien.ac.at \vspace*{8mm} }


\maketitle\abstracts{
We analyze the effect of primordial density perturbations      
on the cosmic QCD phase transition. According to our results   
hadron bubbles nucleate at the cold perturbations.             
We call this mechanism inhomogeneous nucleation.               
We find the typical 
distance between bubble centers to be a few meters. This exceeds 
the estimates from homogeneous nucleation by two orders of magnitude. 
The resulting baryon inhomogeneities may affect primordial nucleosynthesis. 
}

\clearpage

\pagestyle{plain}


The order of the QCD transition and the values of its parameters
are still under debate. Nevertheless there are 
indications from lattice QCD calculations. 
Quenched QCD (no dynamical quarks) shows a 
first-order phase transition with a small latent heat, compared to the 
bag model, and a small surface tension, compared to dimensional arguments 
\cite{Iwasaki}. We assume that the QCD transition is of first order and 
that the values from quenched lattice QCD (scaled appropriately by the 
number of degrees of freedom) are typical for the physical QCD transition.
Based on these values and homogeneous bubble nucleation a small supercooling, 
$\Delta_{\rm sc} \equiv 1- T_{\rm f}/T_{\rm c} \sim 10^{-4}$, and a tiny 
bubble nucleation distance, $d_{\rm nuc} \sim 1$ cm, 
would    
follow \cite{Ignatius}. 
The actual nucleation temperature is denoted by $T_{\rm f}$,
and the thermodynamic transition temperature by                 
$T_c \approx 150$~MeV.                                          
 
We argue~\cite{IS}                                              
that the assumption of homogeneous nucleation is violated in the 
early Universe by the inevitable density perturbations from inflation 
or from other seeds for structure formation. Those 
fluctuations in density and temperature have been measured by COBE 
\cite{Bennett} to have an amplitude of $\delta T/T \sim 10^{-5}$. 
The effect of the QCD transition on density perturbations \cite{SSW,SSW2}
and gravitational waves \cite{Schwarz} has been studied previously, while 
we investigate the effect of the density perturbations on the QCD phase 
transition here.

First-order phase transitions normally proceed via nucleation of bubbles 
of the new phase. When the temperature is spatially uniform and no 
significant impurities are present, the mechanism is homogeneous nucleation. 
The probability to nucleate a bubble of the new phase per time and volume
is approximated by $\Gamma \approx T_c^4 \exp[-S(T)]$.
The nucleation action $S$ is the free energy difference of the system 
with and without the nucleating bubble, divided by the temperature. 

Nucleation is a very rapid process, compared with the extremely slow 
cooling of the Universe. The duration of the nucleation period, 
$\Delta t_{\rm nuc}$, is found to be \cite{Fuller,Enqvist} 
\begin{equation}
  \Delta t_{\rm nuc} = 
  - \frac{ \pi^{1/3} }{{\rm d}S / {\rm d}t \left|_{t_f} \right.} .
\end{equation} 
The time $t_f$ is defined as the moment when the fraction of space where
nucleations still continue equals $1/e$. The heat flow preceding the 
deflagration fronts reheats the rest of the Universe. We denote 
by $v_{\rm heat}$ the effective speed by which released latent heat 
propagates in sufficient amounts to shut down nucleations. In practice, 
$v_{\rm def} < v_{\rm heat} < c_s$, where $v_{\rm def}$ is the velocity 
of the deflagration front and $c_s$ is the sound speed \cite{Kurki-Suonio}.
In the unlikely case of detonations $v_{\rm heat}$ should be replaced by the 
velocity of the phase boundary in all expressions that follow.

The mean distance between nucleation centers, measured immediately after 
the transition completed, is
\begin{equation}
  d_{\rm nuc,hom} = 2 v_{\rm heat} \Delta t_{\rm nuc}
  \label{dnuceq}.
\end{equation}
This nucleation distance sets the spatial scale for baryon number 
inhomogeneities.

Lattice simulations
imply that in real-world 
QCD the energy density must change very rapidly in a narrow temperature 
interval. This can be seen from the microscopic sound speed in the quark 
phase, $c_s \equiv (\partial p/\partial \varepsilon)_{\rm S}^{1/2}$.
Lattice QCD indicates that $3 c_{\rm s}^2(T_{\rm c}) = {\cal O}(0.1)$
\cite{Latticecs2}. Thus, the cosmological time-temperature relation
is strongly modified already before the nucleations, due to
\begin{equation}
  \frac{ {\rm d}T }{ {\rm d}t } = - 3 c_{\rm s}^2 \frac{T}{t_{\rm H}},
  \label{Tteq}
\end{equation}
where $t_H  \equiv  1/H = (3 M_{\rm pl}^2/8\pi\varepsilon_{\rm q})^{1/2}$
with $\varepsilon_{\rm q}$ being the energy density in the quark phase. 
This behavior of the sound speed increases the nucleation distance because 
of the proportionality $\Delta t_{\rm nuc}\propto 1/[3 c_{\rm s}^2 (T_f)]$ 
\cite{Ignatius}.

In the thin-wall approximation the nucleation action has the following 
explicit expression:
\begin{equation}
\label{C}
  S(T)  =  \frac{ C^2 }{(1-T/T_c)^2}, \,\,\,\,\,\,
  C  \equiv  4\sqrt{\frac{\pi}{3}} \frac{\sigma^{3/2}}{l\sqrt{T_c}}\ ,
\end{equation}
for small supercooling.
Assuming further that $c_{\rm s}$ does not change very much during 
supercooling, the following relation holds for the supercooling and
nucleation scales:
\begin{equation}
  \frac{\Delta t_{\rm sc}}{\Delta t_{\rm nuc}} = 
  \frac{\Delta_{\rm sc}}{\Delta_{\rm nuc}} = \frac{2}{\pi^{1/3}} \bar{S}
  \label{Seq}. 
\end{equation}
Here we denote by $\Delta$ a relative (dimensionless) temperature interval
and by $\Delta t$ a dimensionful time interval. $\bar{S} \equiv S(T_f)$ is 
the critical nucleation action, $\bar{S} = {\cal O}(100)$.

Surface tension and latent heat are provided by lattice
simulations with quenched QCD only, giving the values $\sigma = 0.015
T_c^3$, $l = 1.4 T_c^4$~\cite{Iwasaki}.
Scaling the latent heat for the physical QCD 
leads us to take $l = 3 T_c^4$.

With these values for the latent heat and surface tension, the amount of
supercooling is $\Delta_{\rm sc} = 2.3 \times 10^{-4}$. 
{}From Eq.~(\ref{Seq}) it follows that 
$\Delta_{\rm nuc} = 1.5 \times 10^{-6}$. Substituting 
$3 c_{\rm s}^2 = 0.1$ into Eq.~(\ref{Tteq}), we find
$\Delta t_{\rm nuc} = 1.5 \times 10^{-5} t_{\rm H}$ for the duration of the
nucleation period. The nucleation distance depends on the unknown
velocity $v_{\rm heat}$ in Eq.~(\ref{dnuceq}). With the value 0.1
for $v_{\rm heat}$, the nucleation distance $d_{\rm nuc,hom}$ would have 
the value $2.9 \times 10^{-6} d_{\rm H}$. 
One should take these values with caution, 
due to large uncertainties in $l$ and $\sigma$. As our reference set of 
parameters, we take: $\Delta_{\rm sc} = 10^{-4}$, $\Delta_{\rm nuc} = 
10^{-6}$, $\Delta t_{\rm nuc} = 10^{-5} t_H$.

In the real Universe the local temperature of the radiation 
fluid fluctuates. We decompose the local temperature $T(t,{\bf x})$ 
into the mean temperature $\bar{T}(t)$ and the perturbation 
$\delta T(t,{\bf x})$. The temperature contrast is denoted by $\Delta
\equiv \delta T/ \bar{T}$. On subhorizon scales in the radiation dominated 
epoch, each Fourier coefficient $\Delta(t,k)$ oscillates with constant 
amplitude, which we denote by $\Delta_T(k)$. Inflation predicts a Gaussian 
distribution,
\begin{equation}
\label{pd}
p(\Delta){\rm d}\Delta = {1\over \sqrt{2\pi}\Delta_T^{\rm rms}}
\exp\left( - \frac12 {\Delta^2\over (\Delta_T^{\rm rms})^2}\right) 
{\rm d}\Delta \ .
\end{equation}
We find \cite{norm} for the COBE normalized \cite{Bennett} rms temperature 
fluctuation of the radiation fluid (not of cold dark matter)  
$\Delta_T^{\rm rms} = 1.0 \times 10^{-4}$ for a primordial Harrison-Zel'dovich
spectrum. The change of the equation of state prior to the QCD transition  
modifies the temperature-energy density relation, 
$\Delta = c_s^2 \delta\varepsilon /(\varepsilon + p)$. 
We may neglect the pressure $p$ 
near the critical temperature since $p \ll \varepsilon_{\rm q}$ at $T_{\rm c}$. 
On the other hand the drop of the sound speed enhances the amplitude of 
the density fluctuations proportional to $c_s^{-1/2}$ \cite{SSW2}. Putting all
those effects together and allowing for a tilt in the power spectrum, 
the COBE normalized rms temperature fluctuation reads
\begin{equation}
\Delta_T^{\rm rms} \approx 10^{-4} (3 c_s^2)^{3/4} 
\left({k\over k_0}\right)^{(n-1)/2} \ ,
\end{equation}
where $k_0 = (aH)_0$. For a Harrison-Zel'dovich spectrum ($n=1$) and
$3 c_s^2 = 0.1$, we find $\Delta_T^{\rm rms} \approx 2 \times 10^{-5}$. 

A small scale cut-off in the spectrum of primordial temperature fluctuations
comes from collisional damping by neutrinos \cite{Weinberg,SSW2}. The 
interaction rate of neutrinos is $\sim G_{\rm F}^2 T^5$. This has to be 
compared with the angular frequency $c_s k_{\rm ph}$ of the acoustic
oscillations. At the QCD transition neutrinos travel freely on scales 
$l_{\nu} \approx 4 \times 10^{-6} d_{\rm H}$. Fluctuations below the 
diffusion scale of neutrinos are washed out,
\begin{equation}
\label{diff}
l_{\rm diff} = \left[\int^{t_c}_0\!\! l_{\nu}(\bar t) {\rm d} \bar t
\right]^{\frac{1}{2}}\!\! 
\approx 7\!\times\! 10^{-4} d_{\rm H} \label{nueq}\, .
\end{equation}
In Ref.~\cite{SSW2} the damping scale from collisional damping by neutrinos
has been calculated to be $k^{\rm ph}_{\nu} = 10^4 H$ at $T=150$ MeV. The 
estimate (\ref{diff}) is consistent with this damping scale.  
We assume $l_{\rm smooth} = 10^{-4} d_{\rm H}$. 
The compression timescale for a homogeneous volume $\sim l_{\rm smooth}^3$
is $\delta t = \pi l_{\rm smooth}/c_s \sim 10^{-3} t_{\rm H}$. Since 
$\delta t \gg \Delta t_{\rm nuc}$ the temperature fluctuations are 
frozen with respect to the time scale of nucleations. As long as 
$l_{\rm smooth}$ exceeds the Fermi scale homogeneous 
bubble nucleation applies within these small homogeneous volumes.
This is a crucial difference to the scenario of heterogeneous nucleation 
\cite{Christiansen}, where bubbles nucleate at ad hoc impurities.

Let us now investigate bubble nucleation in a Universe with spatially
inhomogeneous temperature distribution. 
Bubble nucleation effectively takes place while the 
temperature drops by the tiny amount $\Delta_{\rm nuc}$. To determine the 
mechanism of nucleation, we compare $\Delta_{\rm nuc}$ with the rms 
temperature fluctuation $\Delta_T^{\rm rms}$:

1. If $\Delta_T^{\rm rms} < \Delta_{\rm nuc}$, the probability to
      nucleate a bubble at a given time is {\em homogeneous} in space. This is
      the case of homogeneous nucleation. 

2. If $\Delta_T^{\rm rms} > \Delta_{\rm nuc}$, the probability to
      nucleate a bubble at a given time is {\em inhomogeneous} in space. 
      We call this inhomogeneous nucleation.

\begin{figure}[t]
\begin{center}
\epsfig{figure=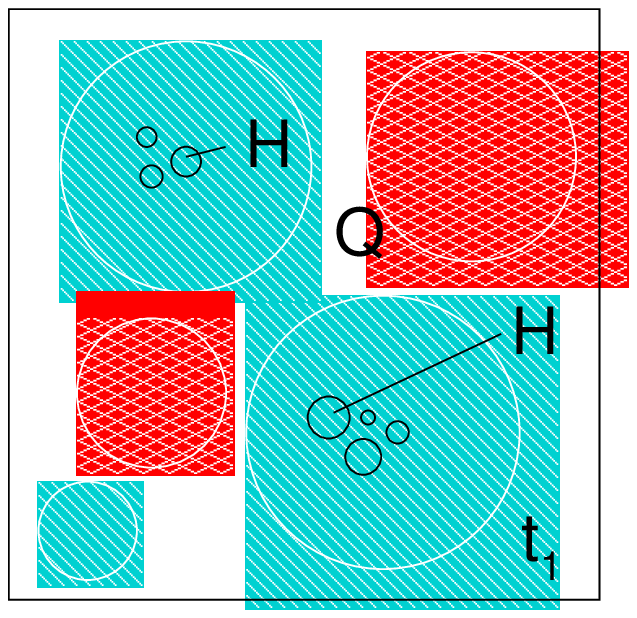,width=0.325\linewidth}          
\epsfig{figure=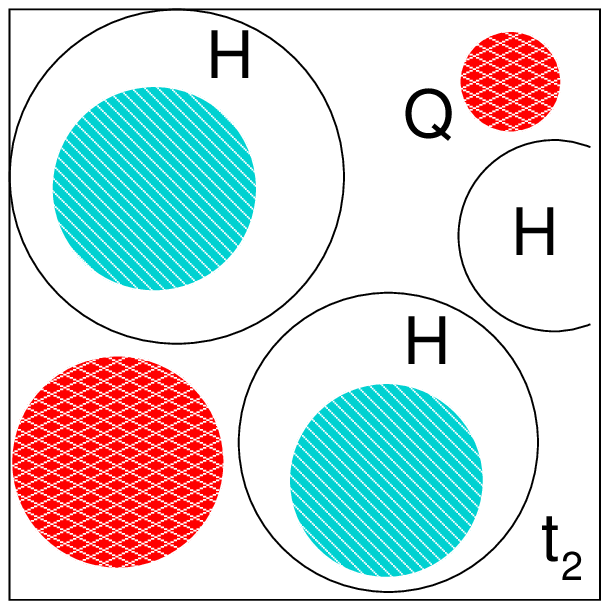,width=0.325\linewidth}          
\epsfig{figure=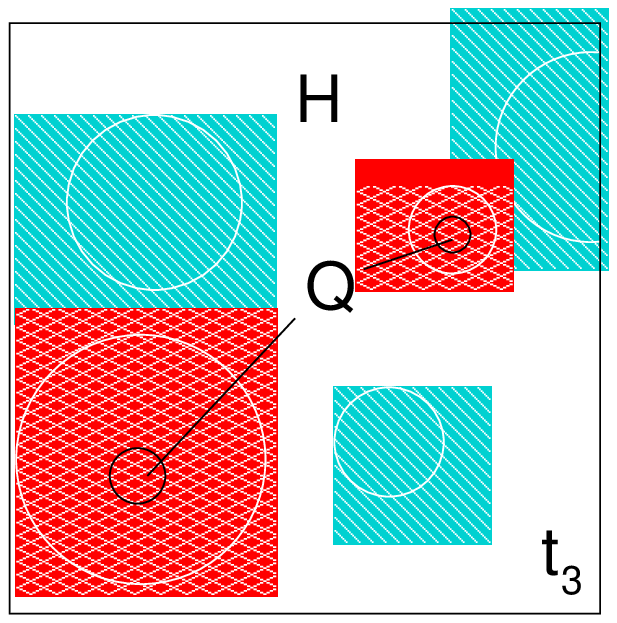,width=0.325\linewidth}          
\end{center}
\caption{\label{fig1}
Sketch of a first-order QCD transition in the inhomogeneous Universe. 
At $t_1$ the first hadronic bubbles (H) nucleate at the coldest spots
(light gray), while most of the Universe remains in the quark phase (Q). 
At $t_2$ the bubbles inside the cold spots have merged and have grown 
to bubbles as large as the temperature fluctuation scale. At $t_3$ the 
transition is almost finished. The last quark droplets are found in the 
hottest spots (dark gray).}
\end{figure}
The quenched lattice QCD data and a COBE normalized flat spectrum 
lead to 
the values $\Delta_{\rm nuc} \sim 10^{-6}$ and $\Delta_T^{\rm rms} \sim 
10^{-5}$. We conclude that the cosmological QCD transition may proceed
via inhomogeneous nucleation. A sketch of inhomogeneous nucleation is shown 
in Fig.~\ref{fig1}. The basic idea is that temperature inhomogeneities 
determine the location of bubble nucleation. Bubbles nucleate first in the 
cold regions.

The temperature change at a given point is governed by the Hubble 
expansion and by the temperature fluctuations. For the fastest changing 
fluctuations, with angular frequency $c_s/l_{\rm smooth}$, we find
\begin{equation}
\label{coolingrate}
{{\rm d}T(t,{\bf x})\over {\rm d} t} = {\bar T\over t_{\rm H}}
\left[- 3 c_s^2 + 
{\cal O}\left(\Delta_T {t_{\rm H}\over \delta t}\right)\right]\ .
\end{equation}
The Hubble expansion is the dominant contribution, as typical values
are $3 c_s^2 = 0.1$ from quenched lattice QCD and $\Delta_T^{\rm rms}
t_{\rm H}/\delta t \approx 0.01$ from the discussion above. 
This means that the local temperature does never increase, except by 
the released latent heat during bubble growth.

To gain some insight in the physics of inhomogeneous nucleation,
let us first inspect a simplified case. We have some randomly 
distributed cold spheres of diameter $l_{\rm smooth}$ with equal and uniform 
temperature, which is by the amount $\Delta_T^{\rm rms}T_c$ smaller than 
the again uniform temperature in the rest of the Universe. When the 
temperature in the cold spots has dropped to $T_{\rm f}$,
homogeneous nucleation takes place in them. Due to the
Hubble expansion the rest of the Universe would need the time
$\Delta t_{\rm cool} = t_H \Delta^{\rm rms}_T  / 3 c_s^2$          
to cool down to $T_{\rm f}$. Inside each cold spot there is a large number 
of tiny hadron bubbles, assumed to grow as deflagrations. 
They merge within $\Delta t_{\rm cool}$ if  
$\Delta_{\rm nuc} < (v_{\rm def}/v_{\rm heat}) \Delta^{\rm rms}_T$.
This condition should be clearly 
fulfilled for our reference set of parameters. Thus the cold spots have fully 
been transformed into the hadron phase while the rest of the Universe 
still is in the quark phase. The latent heat released in a cold spot 
propagates in all directions, which provides the length scale
\begin{equation}
\label{lheat}
  l_{\rm heat} \equiv 2 v_{\rm heat} \Delta t_{\rm cool}. 
\end{equation}
If the typical distance from the boundary of a cold spot to the boundary
of a neighboring cold spot is less than 
$l_{\rm heat}$, then no hadronic bubbles can nucleate in the intervening 
space. In this case the nucleation process is totally dominated by the 
cold spots, and the average distance between their centers gives the 
spatial scale for the resulting inhomogeneities. 
In the following analysis for a more realistic scenario we 
concentrate in this case, $l_{\rm heat} > l_{\rm smooth}$.

The real Universe consists of smooth patches of typical linear size 
$l_{\rm smooth}$, their temperatures given by the distribution~(\ref{pd}).
As discussed above, the merging of tiny bubbles within a cold spot
can here be treated as an instantaneous process. 
The fraction of space that is not reheated by the released latent heat 
(and not transformed to hadron phase), is given at time $t$ by
\begin{equation}
\label{fihn}
f(t) \approx 1 - \int_0^{t}\Gamma_{\rm ihn}(t') V(t,t') {\rm d}t',
\end{equation}
where we neglect overlap and merging of heat fronts. At time $t$ heat, 
coming from a cold spot which was transformed into hadron phase at time 
$t'$, occupies the volume
$V(t,t') = (4\pi/3) [l_{\rm smooth}/2 + v_{\rm heat}(t-t')]^3$.
The other factor in Eq.~(\ref{fihn}), $\Gamma_{\rm ihn}$, is the volume 
fraction converted into the new phase, per physical time and volume as a 
function of the mean temperature $T=\bar{T}(t)$. $\Gamma_{\rm ihn}$ is 
proportional to the fraction of space for which temperature is in the 
interval $[T_{\rm f}, T_f(1 + d\Delta)]$. This fraction of space is given 
by Eq.~(\ref{pd}) with $\Delta = T_{\rm f}/T - 1$. Rewriting ${\rm d}\Delta$ 
by means of Eq.~(\ref{Tteq}) leads to the expression
\begin{equation}
\label{gammai}
\Gamma_{\rm ihn} = 3 c_s^2{T_f\over T}{1\over t_{\rm H} {\cal V}_{\rm smooth}}
p(\Delta = \frac{T_{\rm f}}{T} - 1) , 
\end{equation}
where the relevant physical volume is ${\cal V}_{\rm smooth} = 
(4\pi/3) (l_{\rm smooth}/2)^3.$

The end of the nucleation period, $t_{\rm ihn}$, is defined through 
the condition $f(t_{\rm ihn}) = 0$. We 
introduce the variables $N \equiv (1 - T_{\rm f}/T)/\Delta_T^{\rm rms}$ and 
${\cal N} \equiv N(t_{\rm ihn})$. Since $c_s$ may be assumed to be constant 
during the tiny temperature interval where nucleations actually take place, 
we find from Eq.~(\ref{Tteq}): $1 - t/t_{\rm ihn} \approx 2/(3c_s^2) 
\Delta_T^{\rm rms} (N - {\cal N})$. Putting everything together we determine 
${\cal N}$ from
\begin{equation}
{l_{\rm heat}^3\over l_{\rm smooth}^3} \int_{\cal N}^\infty {\rm d} N 
\frac{e^{-\frac12 N^2}}{\sqrt{2\pi}} 
\left(\frac{l_{\rm smooth}}{l_{\rm heat}} + N - {\cal N}\right)^3 = 1 .
\end{equation} 
The expression                                           
$l_{\rm heat}/l_{\rm smooth} =                           
2 v_{\rm heat}(3 c_s^2)^{-1/4} (k/k_0)^{(n-1)/2}$        
is valid for the COBE normalized spectrum.               
For $l_{\rm heat}/l_{\rm smooth} = 1, 2, 5, 10$ we find 
${\cal N} \approx 0.8, 1.4, 2.1, 2.6$, respectively.

The effective nucleation distance in inhomogeneous nucleation is
defined from the number density of 
those cold spots that acted as nucleation centers, 
$d_{\rm nuc,ihn} \equiv n^{- 1/3}$. 
We find
\begin{eqnarray}
d_{\rm nuc,ihn} & \approx &
\left[\int_0^{t_{\rm ihn}} \Gamma_{\rm ihn}(t) {\rm d} t\right]^{-1/3} \\
& = & 
[\frac{3}{\pi} (1 - {\rm erf}({\cal N}/\sqrt{2})]^{-1/3} \, l_{\rm smooth}. 
\end{eqnarray}
With the above values $l_{\rm heat}/l_{\rm smooth} = 1, 2, 5, 10$ we get 
$ d_{\rm nuc,ihn} = 1.4, 1.8, 3.0,
$ $            
4.8 \times l_{\rm smooth}$, where
$l_{\rm smooth} \approx 1 \mbox{\ m}$. 

For a COBE normalized spectrum without any tilt and with 
a tilt of $n-1 = 0.2$
(where $(k_{\rm smooth}/k_0)^{0.1} \approx 25$),
together with $3c_s^2 = 0.1$ and $v_{\rm heat} = 0.1$, we find the estimate 
$l_{\rm heat}/l_{\rm smooth} \approx 0.4$ and 9, correspondingly. 
Notice that the values of $v_{\rm heat}$ and $3 c_s^2$ are 
in principle unknown.
Anyway, we can conclude that the case 
$l_{\rm heat} > l_{\rm smooth}$           
is a realistic possibility. 

With $2 v_{\rm heat}(3c_s^2)^{-1/4} (10^{-4}d_{\rm H}/l_{\rm smooth}) < 1$
and without positive tilt
we are in the region $l_{\rm heat} < l_{\rm smooth}$, where the geometry is 
more complicated and the above quantitative analysis does not apply.
In this situation nucleations take place in 
the most common cold spots (${\cal N} 
\sim 1$), which are very close to each other. We expect a structure of 
interconnected baryon-depleted and baryon-enriched layers with typical surface
$l_{\rm smooth}^2$ and thickness 
$l_{\rm def} \equiv v_{\rm def}\Delta t_{\rm cool}$. 
In between $d_{\rm nuc,hom}$ 
would be the relevant length scale of inhomogeneities. 
An accurate analysis of this case requires 
computer simulations, which is beyond the scope of the present work. However,
it is clear that the result will be different compared with 
homogeneous nucleation. 

We emphasize that inhomogeneous and heterogeneous
nucleation~\cite{Christiansen} are genuinely different mechanisms, although 
they give the same typical scale of a few meters by chance. 
If latent heat and surface tension of QCD would turn out to reduce  
$\Delta_{\rm sc}$ to, e.g., $10^{-6}$, instead of $10^{-4}$,
the maximal heterogeneous nucleation distance would fall to the 
centimeter scale, whereas on the distance in 
inhomogeneous nucleation this would have no effect.

We have shown that inhomogeneous nucleation during the QCD transition 
can give rise to an inhomogeneity scale exceeding the proton diffusion
scale 
(2~m at 150~MeV).                            
The resulting baryon inhomogeneities could 
provide inhomogeneous initial 
conditions for nucleosynthesis. Observable deviations from the element 
abundances predicted by homogeneous nucleosynthesis seem to be 
possible in that 
case~\cite{MathewsKainulainen}.              

In conclusion, we found that inhomogeneous nucleation leads to 
nucleation distances that exceed by two orders of magnitude 
estimates based on homogeneous nucleation. We emphasize that this new 
effect appears for the (today) most probable range of cosmological and 
QCD parameters.

\section*{Acknowledgments}
We acknowledge Willy the Cowboy for valuable encouragement. We thank 
K.~Jedamzik, H.~Kurki-Suonio, J.~Madsen, and K.~Rummukainen for discussions. 
J.I.~would like to thank the Academy of Finland and
D.J.S.~the Austrian Academy of Sciences for financial support.

\end{document}